\documentclass[preprint,a4]{acm_proc_article-sp}
\usepackage{epsfig}
\usepackage[square,comma,numbers,sort&compress]{natbib}
\usepackage{multirow}

\newcommand{\UPLB}{University of the Philippines Los Ba\~{n}os}
\newcommand{\CROWDSIZE}{c}
\newcommand{\EXITWIDTH}{w}
\newcommand{\CORRWIDTH}{W}
\newcommand{\TIMEARCH}{T}
\newcommand{\SIZEARCH}{S}
\newcommand{\PERSON}{\mathcal P}

\begin{document}
\title{A Study on the Effect of Exit Widths and Crowd Sizes in the Formation of Arch in Clogged Crowds}
\numberofauthors{1}
\author{
\alignauthor Francisco Enrique Vicente G. Castro and Jaderick P. Pabico\\
   \affaddr{Institute of Computer Science}\\
   \affaddr{\UPLB}\\
   \affaddr{College 4031, Laguna}
   \email{\{fevgcastro,jppabico\}@uplb.edu.ph}
}
\date{}
\maketitle

\begin{abstract}
The arching phenomenon is an emergent pattern formed by a $\CROWDSIZE$-sized crowd of intelligent, goal-oriented, autonomous, heterogeneous individuals moving towards a $\EXITWIDTH$-wide exit along a long $\CORRWIDTH$-wide corridor, where $\CORRWIDTH>\EXITWIDTH$. We collected empirical data from microsimulations to identify the combination effects of~$\CROWDSIZE$ and~$\EXITWIDTH$ to the time~$\TIMEARCH$ of the onset of and the size~$\SIZEARCH$ of the formation of the arch. The arch takes on the form of the perimeter of a half ellipse halved along the minor axis. We measured the~$\SIZEARCH$ with respect to the lengths of the major~$M$ and minor~$m$ axes of the ellipse, respectively. The mathematical description of the formation of this phenomenon will be an important information in the design of walkways to control and easily direct the flow of large crowds, especially during panic egress conditions.
\end{abstract}

\section{Introduction}\label{sec:intro}
A crowd is a large group of humans, usually numbering by the hundreds to tens of thousands, who are in the same physical environment, such as a hallway, sharing a common purpose, such as exiting through a sole door along the hallway. Although the individuals in the crowd are sharing a common purpose, they may act differently in a crowd than when they are by themselves~\citep{helbing00}. The general movement of a crowd is affected by a number of different individuals who are moving independently from each other, but sharing a common environment, and aiming to reach their respective destinations. The understanding of crowd movement is important to planning and improving shared public places, not only to effectively and efficiently facilitate the comfortable movement of individuals, but also to guarantee the individual safety, especially under conditions of danger when quick and orderly evacuation of a mass of individuals is desired.

The patterns of behavior of individuals in a crowd is affected directly by the interactions and influences between the individuals and their environment. Because data from real evacuation is hard to measure and conducting experiments on humans has ethical questions, the characterization of the dynamic aspects of evacuation processes have been confined by experts (engineers, sociologists, and scientists alike) only through simulation and modeling. Because of this, several computer models have been proposed and utilized which aided in the understanding of crowd behaviors under normal, controlled egress, and panic situations. Of course, these models have been separately evaluated either by comparing the existing descriptive real-world data usually compiled by social scientists with the descriptive behavior of the model outputs, or by comparing the quantitative and qualitative outputs of the model with that of another model.  One among the many models of crowd dynamics is the social force model (SFM)~\citep{helbing00} which describes the motion of an individual as affected by both physical and psychological forces. SFM has been a popular model among researchers and planners because it was successful in describing phenomena observed in the real world, such as the ``faster-is-slower'' in escape panic and the ``arching'' phenomenon observed as a side-effect to ``clogging'' on exit ways. However, when using SFM for crowds under low crowd density conditions, empirical results show that the simulated individuals do not behave as expected. In other words, the simulated individuals act irrationally rather than rational ones, as we expect of humans under normal conditions. For example, the simulated individuals go directly and repeatedly towards an object that blocks an exit way, instead of finding a way to go around the object and exit the hallway. This happens because the desired direction of the simulated individual is always towards the exit points, even in the presence of obstacles between its current position and the exit points.

The SFM has been used extensively by many researchers~\citep{helbing95,helbing00,helbing01,helbing06,henein05} because of its capability to simulate the often observed human crowd phenomena such as flocking and bidirectional lane formation, among others. However, the simulated individuals under SFM do not exhibit some of the common human behaviors observed in a large crowd. These behaviors are:
\begin{enumerate}
\item {\em Imitation}. According to~\citet{festinger54}, there is a drive in a person towards its own self evaluation. This drive forces the person to belong to groups so that he can be associated with others. Thus, in a crowd of moving people, the individuals tend to move into groups whose members they believe would have the same opinion as theirs.
\item {\em Contagion}. Using the imitation behavior, people will also tend to ``adopt'' the behavior of others in the same group. Thus, if a group member senses that there is a discrepancy in a group with respect to behavior, opinions or abilities, that same member will reduce the discrepancy by adopting the most common behavior, opinion, or ability~\citep{festinger54}.
\end{enumerate}
Because of these, researchers~\citep{fridman10} recently proposed the social comparison theory (SCT) that algorithmically mimics the above-mentioned behaviors and was introduced as an alternative improvement to SFM. The key idea of SCT is that humans who lack the objective means to evaluate their current state, such as when in confusion, would compare themselves to others that they perceive are similar to themselves (i.e., both imitation and contagion behaviors).

In our previous work~\citep{castro12}, we used the SCT to design an artificial individual whose behavior resembles that of a human when walking (or traveling) within a crowd. We included in the same model a ``field of desire'' to allow the simulated individual to change its desired direction of movement voluntarily, based on its perception of the behavior and location of other individuals and obstacles along the hallway. The ``field of desire'' is a $100^\circ$ field of vision facing the forward motion of the simulated individual, which the individual uses to obtain spatial information concerning its surroundings and to choose the desired forward direction towards the exit goal. Our crowd of simulated individuals were able to exhibit the arching, clogging, and bursty exit phenomena usually observed in real crowd situations. The arch exhibited at the edge of the clogged crowd resembles that of a half ellipse, with the major axis parallel with the direction of the crowd flow, while the minor axis is coincident with the axis of the exit width.

We present in this paper the result of the study we conducted to understand the interactive effects of crowd size~$\CROWDSIZE$ and exit width~$\EXITWIDTH$ in the formation of arch in a clogged crowd moving towards one exit door along and in the middle of a hallway. We will show by empirical means that the onset of arching has a negative correlation with the increasing~$\EXITWIDTH$, while its slope tends to grow steeper as the~$\CROWDSIZE$ is increased. We will also show that the profile of the arch resembles a half ellipse whose major axis runs along the length of the hallway, while its minor axis is parallel to the width of the exit door. The major axis grows smaller as the exit width is increased, but grows bigger with the increase of the crowd size. The minor axis, on the other hand, grows bigger either as the crowd size is increased, or as the exit width is increased (of course limited only by the width of the hallway).

\section{Review of Literature}\label{sec:review}

\begin{figure*}[hbt]
\centering\epsfig{file=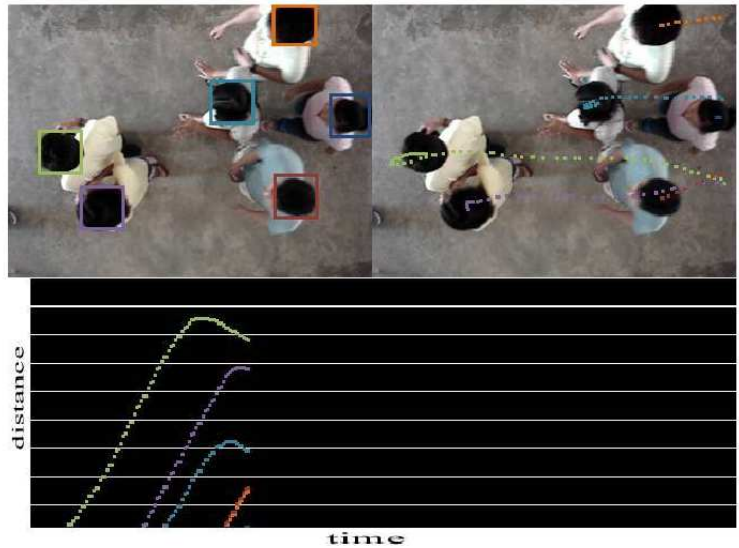, width=100mm}
\caption{An example output frame of the system developed by~\citet{ngoho09}. The upper left part contains the input image of student pedestrians with trackers (bounding boxes). The upper right part contains the trajectory of the pedestrians. The lower part contains the distance vs. time graph of the trajectory. This figure is in color in the electronic copy of this paper printed with permission from~\citet{ngoho09} and the Philippine Society of Information Technology Educators Foundation, Inc. (PSITE).}\label{fig:1}
\end{figure*}

The approaches used by researchers for crowd simulation maybe classified into two categories namely, macroscopic and microscopic. The macroscopic models  are models that describe the dynamics of observable quantities of a crowd motion, such as density and velocity. Examples of such models are the ones used by~\citet{hughes02}, \citet{huang09}, and~\citet{treuille06} wherein they captured the density and velocity of the crowd using partial differential equations, the usual mathematical descriptions of the dynamics of fluid. Similar to what~\citet{ngoho09} did in the past, ~\citet{musse07} proposed a model based on the trajectories of motion captured automatically from video sequences. While~\citet{ngoho09} captured the real-time trajectory of the velocity field (see for example Figure~\ref{fig:1}), \citet{musse07} used an offline method to cluster similar trajectories to extrapolate the velocity field.

Borrowing some ideas in gas-kinematic models, \citet{helbing95} introduced the SFM to simulate pedestrian flows, wherein a self-driven particle interacts through social rules and regulations, moves in its desired speed and direction, and attempts not to collide with obstacles, other particles, and surrounding barriers.  Thus, in order to reach its destination faster, pedestrians take detours even if the taken route is crowded~\citep{helbing01}. However, the choice of detour is dependent on the recent memory of what the traffic was like the last time they took the route, the figure of which was found by~\citet{ganem98} to be polygonal in nature. In agreement with the SFM, ~\citet{weidmann93} observed that, as long as it is not necessary to go faster, such as going down a ramp, a pedestrian prefers to walk with his or her desired speed, which corresponds to the most comfortable walking speed. However, \citet{weidmann93} further observed that pedestrians keep a certain distance from other pedestrians and borders. The distance between the pedestrians decreases as the density of the crowd increases. The pedestrians themselves cause delays and obstructions. \citet{arns93} observed that the motion of the crowd is similar to the motion of gases and fluid, while~\citep{helbing01} suggested that it is similar to granular flow as well. 

\citet{helbing01}, in his extension of the SFM, showed that many aspects of traffic flow can be reflected by self-driven many-particle systems. In this system, he identified the various factors that govern the dynamics of the particles such as the specification of the desired velocities and directions of motion, the geometry of the boundary profiles, the heterogeneity among the particles, and the degree of fluctuations. One such observable pattern is the formation of lanes of uniform walking direction, formed because of the self-organization of the pedestrians~\citep{helbing06}. Aside from the self-organizing behavior of the crowd, obstacles were also observed to both positively and negatively contribute to the flow of the traffic. During escape panic of large crowds, several behavioral phenomena were observed~\citep{helbing00}: build up of pressure, clogging effects at bottlenecks, jamming at room widening areas, faster-is-slower effect, inefficient use of alternative exits, initiation of panics by counter flows, and impatience. It was observed that the main contributing behaviors in these situations is a mixture of individual and grouping behavior.

While macroscopic models describe the dynamics of the collective crowd, microscopic models describe the speed and position of each crowd member in a given time. Historically, \citet{reynolds87} is known to have pioneered the simulation of crowd in this manner. In his work, he simulated a flock of birds where each simulated bird is implemented as an independent individual that navigates according to its local perception of the dynamic environment. In a similar study in the years that follow,~\citet{tu94} simulated schools of artificial fishes using perception to drive each fish's locomotion while being constrained by underwater physics. The then novel programming of the fish's motion and environment resulted into observed real-world school of fishes' behaviors.

Because of the success of microsimulation in exhibiting observed behaviors in animals, researchers in various fields of science (e.g., transportation science, computer gaming, artificial intelligence, and industrial robotics) adopted the programming method to simulate the motion and environment of human pedestrians in a crowd. The main purpose is to aid them in understanding some emergent crowd behaviors so that they can design better and safer walkways, or build realistic gaming elements, environments, and scenarios. During the advancement of the method itself, several enhancements to the model, as well as alternatives, have been suggested. For instance, ~\citet{kapadia09} proposed the use of what they called an ``egocentric affordance field'' to obtain the speed and the direction of each simulated pedestrian. On the other hand, \citet{varas07} and~\citet{perez02} introduced cellular automata (CA) models to direct the motion of the simulated crowd member. Both works located their crowd members at cells of a grid wherein their respective coordinates are updated at discrete time intervals. The implementation of the CA model was so simple that~\citet{teknomo11} implemented both the CA simulation and visualization using only a desktop electronic spreadsheet program. The simplicity of CA prompted other researchers~\citep{marumatsu99, tajima01, gou08} to further improve it using the probabilistic and statistical functions used in lattice-gas models to direct the motion of the simulated member in a cell. The lattice-gas model is known to be a special case of CA.

Although the above models were already successful in simulating observed phenomena in crowds, the behavior of the crowd member as an individual has not been fully integrated into these models. Because of this,~\citet{lerner09} proposed an approach for fitting human behaviors, such as talking on the phone, combing hair, and moving the head, into the simulated individual. In an attempt to simulate the collective behavior of humans in groups, ~\citet{musse01} used a hierarchical crowd organization with autonomous groups to simulate the grouping or flocking behavior in pedestrians. To be able to model and generate low-level behaviors for the simulated individuals, ~\citet{goldstein99} presented non-linear dynamic systems to model the interaction of the simulated individuals to its environment. Such low-level behaviors are real-time target tracking and obstacle avoidance in a highly dynamic environment. Later, the authors~\citep{goldstein01} used a hybrid low-level system integrating the non-linear dynamical systems with kinetic data structures to simulate a complex crowd environment that includes three-dimensional steering, crowding, and flocking within moving and non-moving obstacles.

Recently, established works in the field of social psychology have been successfully incorporated into microsimulation models of the crowd, resulting into a much more realistic simulated collective behavior. Example of works go as far as 1895 such as that of~\citet{lebon1895}, and the newer ones such as that of~\citet{allport24},~\citet{blumer39}, and~\citet{berk74}. In these works, the authors discussed that people in crowds act similarly to one another, usually in a coordinated fashion that when viewed from a distance, it would appear that they are being choreographed by a single entity. However, in reality, the coordinated motion is achieved with little or no verbal communication at all. This somewhat homogeneous motion by the crowd is explained by~\citet{lebon1895} as being directed by two processes, {\em imitation} and {\em contagion} that we discussed earlier in Section~\ref{sec:intro}. The coordinated motion occurs in a ``circular reaction'' process, a type of interstimulation between two individuals~$a$ and~$b$ where the response of~$b$ to the stimuli from~$a$ reproduces a stimulation that is reflected back to~$a$. The reflected stimulation reinforces~$a$'s stimuli in return, thus resulting into a circular reaction process~\citep{blumer39}. Following the adage that ``similar people act in similar ways,'' \citet{allport24} explained that the somewhat homogeneous motion of the crowd is the product of likeminded individuals. The likemindedness of the individuals in the crowd has been given a game-theoretic perspective by~\citet{berk74}. Here, he said that the coordinated behavior of the crowd is consistent with that of an individual applying a minimax strategy.  Given an action~$A$ being performed by the greatest number of individuals in a crowd, an individual~$c$ will see~$A$ as the action with least cost, and thus will perform it also. The individual that selects the action of the majority has to pay the least cost compared to that which follows the minority.

One of the most recent models that incorporates works in social psychology, including that of cognitive sciences, is that of~\citet{fridman10} where they used the ideas of SCT by~\citet{festinger54}. In their work, they proposed an algorithmic framework for SCT which resulted into simulated crowd behavior that is more in tune with observed real-world crowd behavior compared to those techniques that do not incorporate the cognitive models. Because of this, we have decided to start with the SCT, but hybridized it with our own algorithm for mapping the perceived trajectory of an individual's goal, limited by the individual's field of vision (i.e., our ``field of desire'')~\citep{castro12}. Our hybrid model was able to exhibit important crowd behaviors: arching, clogging, and bursty exit rates. We have also observed a new phenomenon we called double arching, which we will discuss in detail in another paper and forum. In this paper, we will explore the interactive roles of~$\CROWDSIZE$ and~$\EXITWIDTH$ to the formation of the arch using the hybridized microsimulation approach in the hope of exploiting the outcomes for designing better and safer walkways under various crowd egress situations.  

\section{Methodology}\label{sec:method}

\subsection{Programming the Individual}
In our approach, each simulated individual was programmed with the following general behaviors which were described originally by~\citet{wooldridge95}, and then later on extended and explained further by~\citet{frankling96,epstein99,torrens04} and~\citet{macal10}:

\begin{enumerate}
\item {\em Autonomy}: The individuals are autonomous units, and thus are not directed by any centralized control. They are capable of processing information and exchanging this information with other individuals in order to make independent decisions. They may freely interact with others in a few given situations, and this freedom does not affect their being autonomous. With this, we say that the individuals are active rather than purely passive (see discussion of ``active'' below). 
\item {\em Heterogeneity}: We believe that individuals should be heterogeneous to allow the microsimulation to develop autonomous individuals. 
\item {\em Active}: Being active means that the individuals influence a simulation independently. The following active features can be identified: 
  \begin{enumerate}
  \item Pro-active or goal-directed
  \item Reactive or perceptive
  \item Bounded rationality
  \item Interactive and communicative
  \end{enumerate}
\end{enumerate}

Aside from the abovementioned general behaviors, we infuse the SCT by following the algorithmic approach of~\citet{fridman10}, which we describe here for clarity. The SCT follows a subset of axioms as follows:

\begin{enumerate}
\item {\em When individuals can not objectively decide what to do next, they compare their current state with that of others.} Each individual $\PERSON_i$ is characterized by a tuple with $k$ attributes $\PERSON = \langle a^\PERSON_1, a^\PERSON_2, \dots, a^\PERSON_k \rangle$, where each attribute $a^i_j$  of individual $\PERSON_i$, $\forall 1\le j \le k$, corresponds to a point in a $k$-dimensional space. Each dimension corresponds to an attribute such as the coordinates of $\PERSON_i$ in two-dimensional space, heading, height, gender, etc.
\item {\em Individuals compare their state to individuals that are similar to them.} We defined a similarity function ${\bf Sim}(\PERSON_i, \PERSON_j)$ which measures the similarity between the $i$th and $j$th individuals ($i\ne j$). The similarity function is given in Equation~\ref{eqn:sim}. For an individual $\PERSON_i$, we selected the individual with the highest similarity.
\item {\em If there are differences in their state with the object of comparison, the individuals take steps to reduce the differences.} If the similarity of the selected individual is below a given value, then $\PERSON_i$ is triggered to do something to reduce the difference.
\end{enumerate}

We measured the similarity between $\PERSON_x$ and $\PERSON_y$ independently along each dimension, such that the similarity in dimension $a_i$ is  a function $S_{a_i}(a_i^x, a_i^y): a_i \times a_i \rightarrow [0, 1]$. A value of 1 indicates full similarity, while a 0 means dissimilarity. For example, one attribute of $\PERSON_x$ is its distance from $\PERSON_y$, thus a value of 1 means they are positioned in the same location, while a 0 means they are far apart.

\begin{equation}
{\bf Sim}(\PERSON_x, \PERSON_y) = \sum_{j=1}^k S_{a_j}(a^x_j, a^y_j) \times w_j\label{eqn:sim}
\end{equation}

\begin{figure*}[hbt]
\centering\epsfig{file=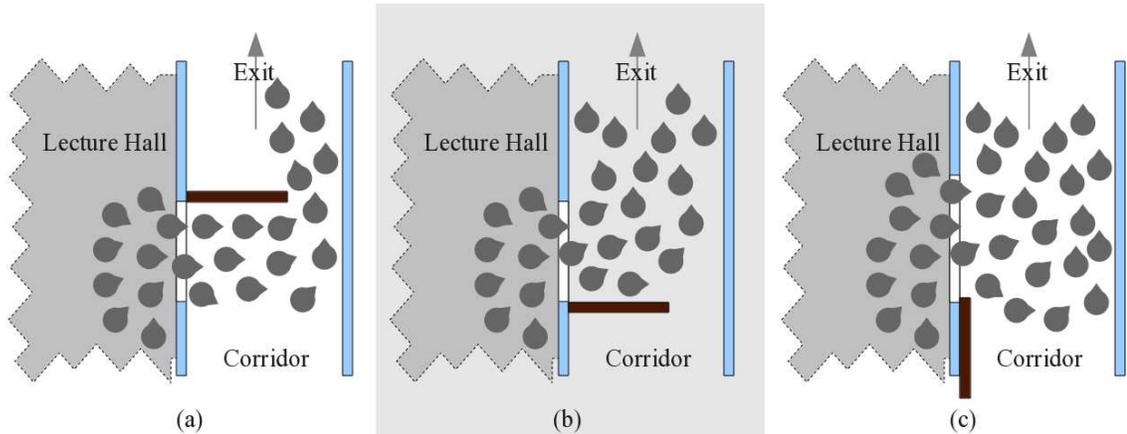, width=150mm}
\caption{Snapshots of example {\em what-if} scenario simulations on the effect of exit door configurations to student egress (students are shown as black circles with pointers to visualize heading) in a very large lecture hall: (a) Current exit door configuration which opens outward but hinders student flow; (b) Effect of moving the door hinge such that the exit door opens behind the student flow; and (c) Effect of increasing the effective width of the exit door by 50\% and replacing the swinging door by a sliding door. This figure is in color in the electronic copy of this paper printed with permission from~\citet{castro13} and the Philippine Society of Information Technology Educators Foundation, Inc. (PSITE).}\label{fig:2}
\end{figure*}

Because our simulated individuals exhibit these behaviors, our simulation approach is more akin to modeling humans and objects in very realistic ways than other modeling approaches, such as those that aggregate mathematical equations to explain the dynamics of pedestrian movement (i.e., macrosimulations). Thus, our simulation can be used with higher confidence to perform what-if scenarios to aid, for example, university administrators and decision makers with regards to management policies, as well as infrastructure development plans, for safer learning environments for the students and constituents (see for example Figure~\ref{fig:2}).

\subsection{The Goal of the Individual}

In a two-dimensional flat world, let $G$ be the goal of $\PERSON_j$ ($\forall j$) with set coordinates $\{(x_1, y_1), (x_2, y_2), \dots, (x_n, y_n)\}$ representing a line segment. This set is usually the location of a fixed-width exit door. Since all $\PERSON$ aim to reach any one of the exit coordinates, a given $\PERSON_i$ will face towards the direction of the nearest exit coordinates from its current coordinates. The $\PERSON_i$ will scan its field of vision for the closest free space and move towards that space with its gaiting speed, which is currently set at one pace per simulation time step. In the future, we wish to vary this gaiting speed depending on the height of~$\PERSON_i$. A free space is a location in the flat world that is not a wall, a~$\PERSON_j$ where $j\ne i$, or any other object. If other individuals are blocking $\PERSON_i$'s space within its field of vision, $\PERSON_i$ stops at its current coordinates. It is possible that the individual might move away from the target exit coordinate if the chosen free space was at the edge of its field of vision. When that happens, the individual will still move to the free space, but after moving, it will redirect its heading towards the possibly new nearest exit coordinates. When the distance of the individual to the nearest exit coordinates is $<1$, it considers itself as already exited and will move to the edge of the simulation world. Until this distance has not been achieved, the individual will just repeat its decision making process as described above.

\subsection{Interactive Effects of Crowd Size and Exit Width on Arching Profile}

We conducted experiments along a fixed-width hallway to determine the interactive effects of~$\CROWDSIZE$ and~$\EXITWIDTH$ to the time~$\TIMEARCH$ of the onset of formation of the arch, as well as the dimension of the arch's profile at~$\TIMEARCH$. The arch profile is a half ellipse, halved across the major axis, and is measured with respect to the length of the minor and major axes. We experimented on five levels of~$\CROWDSIZE$ (200, 300, 350, 400, and 450 individuals) and seven levels of~$\EXITWIDTH$ (1, 3, 5, 7, 9, 11, and 13), resulting into 35 $\CROWDSIZE$--$\EXITWIDTH$ combinations. We replicated each combination three (3)~times to take into consideration random variabilities brought about by external factors that can not be modeled, such as the implementation of the random number generator of our host operating system and other systemic variabilities such as the occurrences of hardware interrupts during simulation that might have affected our measurements. The exit width is defined as the number of individuals in a straight line that can concurrently and comfortably yet tightly fit along the axis of the exit width. Similarly, the fixed-width hallway, set at 19, is also defined as the number of individuals along a straight line that can comfortably yet tightly fit in a straight line parallel to the corridor width.

\section{Results and Discussion}\label{sec:results}

\begin{figure*}
\centering\epsfig{file=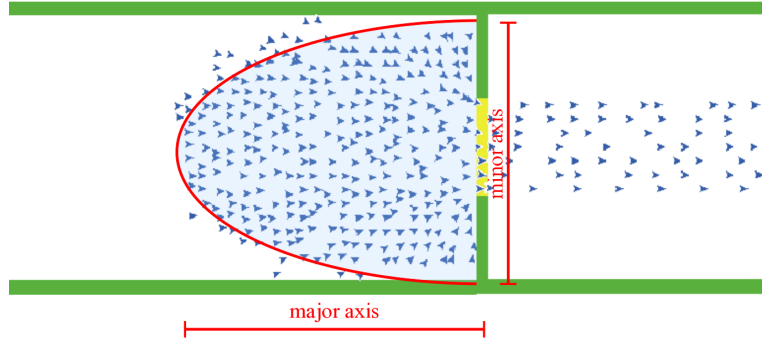, width=100mm}
\caption{Profile of the crowd before the exit door as it forms an arch. This is a snapshot of the microsimulation when the $\CROWDSIZE=400$ at $\EXITWIDTH=7$. Notice that the major axis runs along the axis of the corridor, while the minor axis is parallel with the axis of the exit door.}\label{fig:arching-example}
\end{figure*}

Figure~\ref{fig:arching-example} shows an example profile of a crowd that exhibits arching near the exit door. We see here that the arch resembles a half ellipse with the ellipse's major axis running along the direction of the crowd flow and perpendicular to the exit width, while the minor axis is parallel to the exit width. This result is a confirmation of the results of~\citet{frankling96,epstein99} and~\citet{helbing01}. 

\subsection{Arching Time}

\begin{figure*}
\centering\epsfig{file=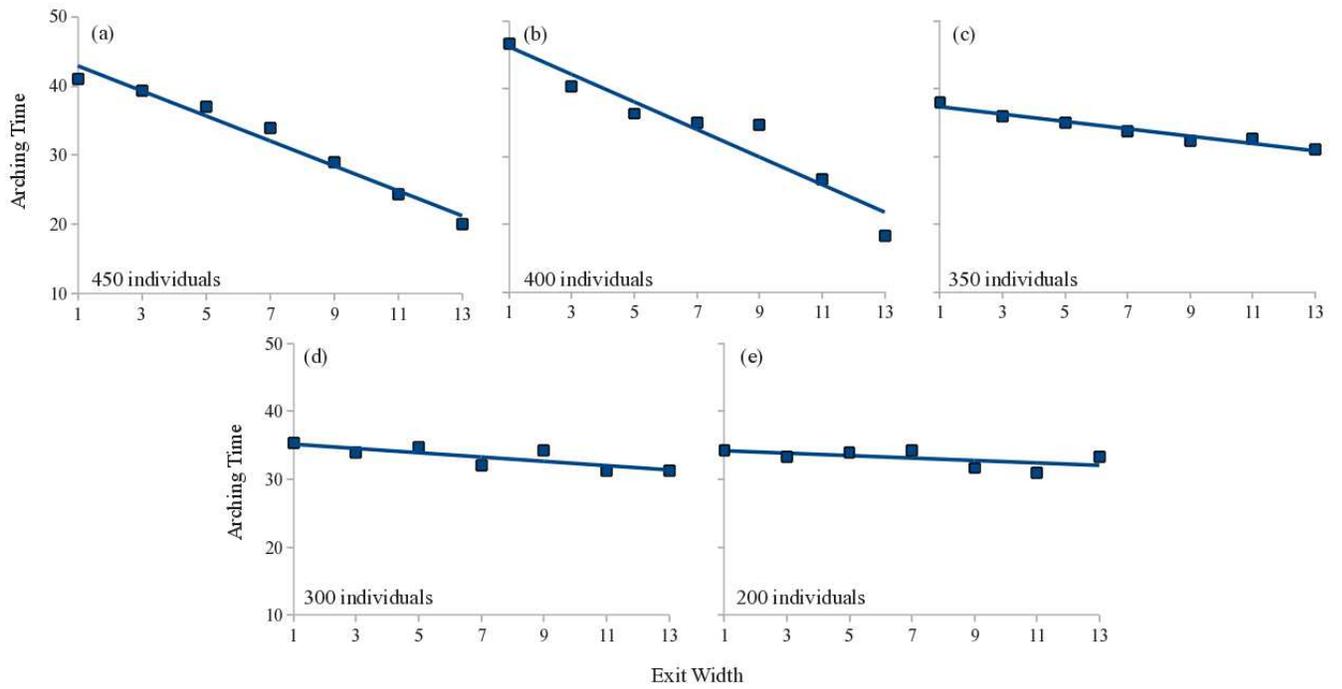, width=7in}
\caption{$\TIMEARCH$ as affected by $\CROWDSIZE$ and $\EXITWIDTH$ combination: (a) $\CROWDSIZE=450$; (b) $\CROWDSIZE=400$; (c) $\CROWDSIZE=350$; (d) $\CROWDSIZE=300$; and (e) $\CROWDSIZE=200$. The blues squares are means of~$\TIMEARCH$ at each $\CROWDSIZE$--$\EXITWIDTH$ combination, averaged across all replicates. The blue lines are the respective regression lines.}\label{fig:arching-time}
\end{figure*}

Figure~\ref{fig:arching-time} shows the response of the onset of arching time~$\TIMEARCH$ to the combination effect of varying~$\CROWDSIZE$ and~$\EXITWIDTH$. We can see here that across different~$\CROWDSIZE$, there is a recurrent pattern that the~$\TIMEARCH$ decreases as the~$\EXITWIDTH$ is increased, but the amount of decrease is not the same between any~$\CROWDSIZE$. To better quantify this observed decrease in~$\TIMEARCH$ at increasing~$\EXITWIDTH$, we conducted a regression analysis between~$\TIMEARCH$ and~$\EXITWIDTH$ for each value of~$\CROWDSIZE$. Equations~\ref{eqn:t450} through~\ref{eqn:t200} show these regressions, respectively for $\CROWDSIZE = \{450, 400, 350, 300, 200\}$. All coefficients are significantly different from zero at a confidence level of $\alpha=0.01$, while all $R^2$ values are significantly different from zero at~$\alpha=0.05$. This means that even though Equations~\ref{eqn:t300} and~\ref{eqn:t200} have low $R^2$ values, these values are non zero. Looking at the respective slopes of the regression, we see here that as the $\CROWDSIZE$ is decreased, the slope gets to zero. This means that at lower $\CROWDSIZE$ values, the onset of the formation of the arch is becoming independent of the~$\EXITWIDTH$. The~$\EXITWIDTH$ has a significant effect on~$\TIMEARCH$ at higher~$\CROWDSIZE$. Generally, however, $\TIMEARCH \propto \frac{1}{\CROWDSIZE \cdot \EXITWIDTH}$.

\begin{eqnarray}
   \TIMEARCH_{450} & = & -1.80 \EXITWIDTH + 44.72, R^2 = 0.97\label{eqn:t450}\\
   \TIMEARCH_{400} & = & -2.04 \EXITWIDTH + 48.25, R^2 = 0.92\label{eqn:t400}\\
   \TIMEARCH_{350} & = & -0.54 \EXITWIDTH + 37.89, R^2 = 0.95\label{eqn:t350}\\
   \TIMEARCH_{300} & = & -0.32 \EXITWIDTH + 35.49, R^2 = 0.66\label{eqn:t300}\\
   \TIMEARCH_{200} & = & -0.18 \EXITWIDTH + 34.39, R^2 = 0.34\label{eqn:t200}
\end{eqnarray}

\subsection{Arch Profile}

\begin{figure*}
\centering\epsfig{file=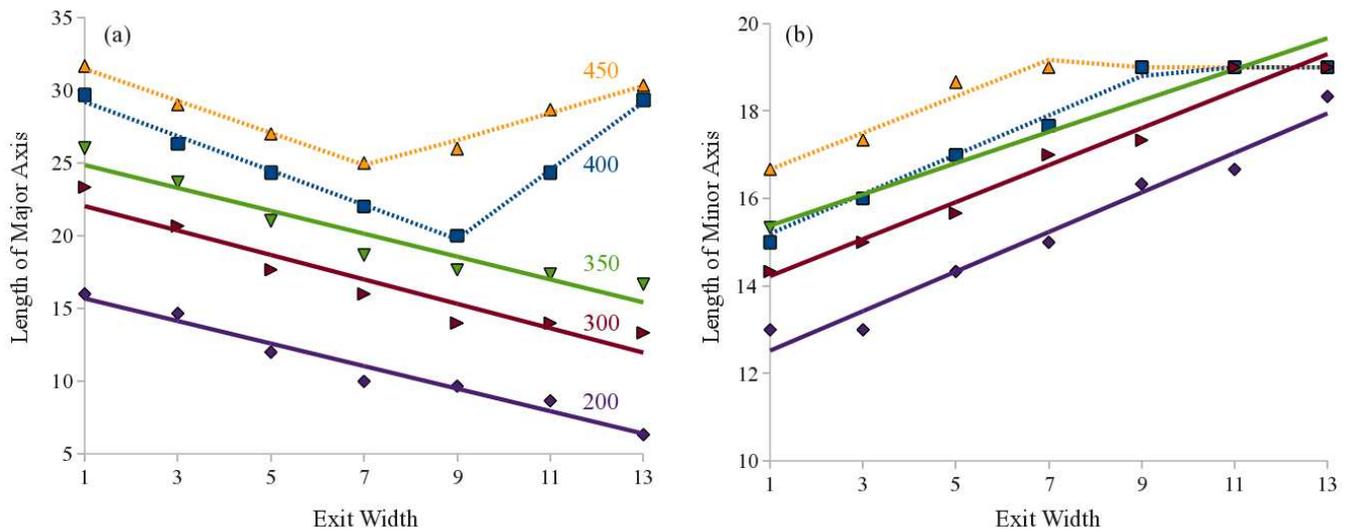, width=7in}
\caption{(a) $M$ and (b)~$m$ as affected by $\CROWDSIZE$ and $\EXITWIDTH$ combination. Colored shapes are means while lines are regression fit at each~$\CROWDSIZE$. Orange/triangle represents $\CROWDSIZE=450$, blue/square $\CROWDSIZE=400$, green/inverted triangle $\CROWDSIZE=350$, red/triangle pointing right $\CROWDSIZE=300$, and purple/diamond $\CROWDSIZE=200$.}\label{fig:arching-profile}
\end{figure*}

Figure~\ref{fig:arching-profile} shows the response of the lengths of the major and minor axes at varying values of~$\CROWDSIZE$ and $\EXITWIDTH$. Except for $\CROWDSIZE$ at 450 and 400 individuals, the length of the major axis decreases as the $\CROWDSIZE$ is decreased, and as the~$\EXITWIDTH$ is increased, while the length of the minor axis decreases as the~$\CROWDSIZE$ is decreased, and increases as the~$\EXITWIDTH$ is increased. Mathematically, $M \propto \frac{\CROWDSIZE}{\EXITWIDTH}$, while $m \propto \CROWDSIZE \cdot \EXITWIDTH$.

\begin{figure*}
\centering\epsfig{file=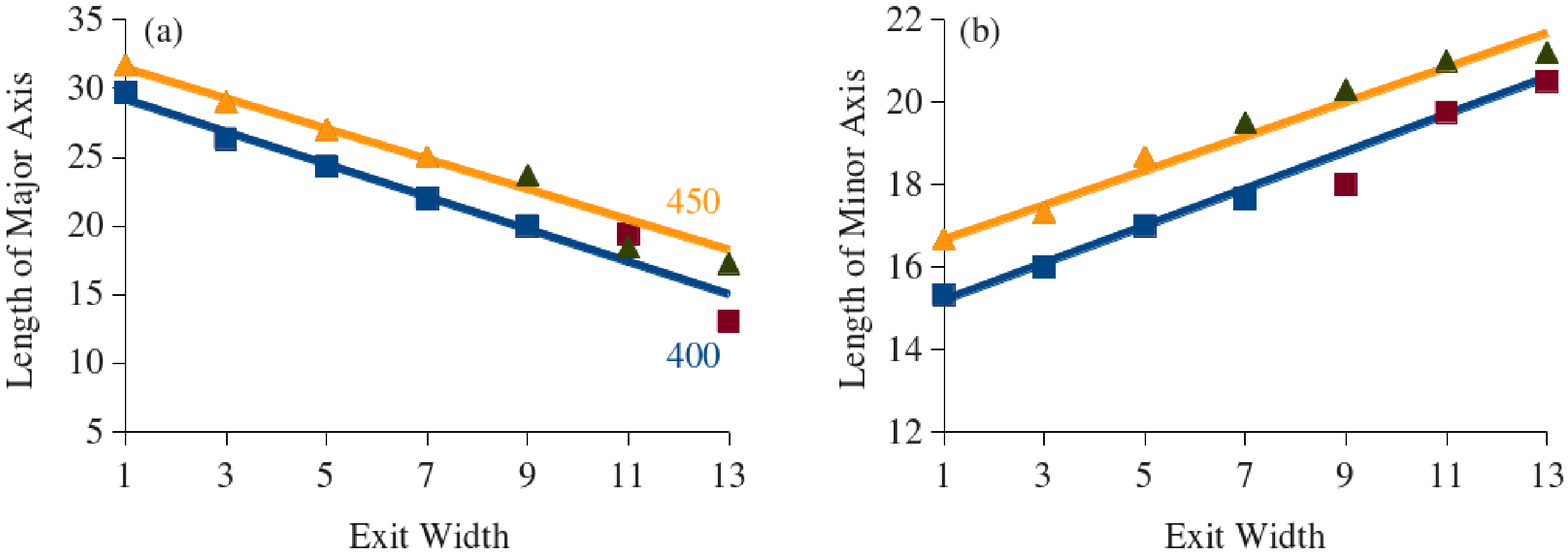, width=7in}
\caption{(a) $M$ and (b)~$m$ as affected by $\CROWDSIZE$ and $\EXITWIDTH$ combination when $\CORRWIDTH=35$.}\label{fig:assert}
\end{figure*}

We observed here from the figure that when the $\EXITWIDTH=9$, the $\CROWDSIZE=400$ plateaued at $m=19$, while with~$M$, the slope becomes positive. Similar behavior was observed for $\CROWDSIZE=450$ but the plateauing and the sudden change of direction of the slope for~$m$ happened at $\EXITWIDTH=7$. At these~$\CROWDSIZE$ values, the individuals have already reached the edge of the corridor that~$m$ cannot expand anymore. Since~$m$ is constrained to 19, the arch must accommodate more individuals, thus translating into the increase in~$M$. However, had the hallway width been way more than 19, the same pattern would have been observed for all~$\CROWDSIZE$. We confirmed this assertion when we rerun the simulation for~$\CROWDSIZE = \{400, 450\}$ and~$\EXITWIDTH = \{9, 11, 13\}$ at the same number of replicates. When we plotted the mean of each $\CROWDSIZE\times\EXITWIDTH$ combination across replicates, they fall reasonably well within the regression line that we obtained earlier, of course within some acceptable variance (i.e., the respective means' $\pm 1 \times$ standard deviations intersect with the regression line, see Figure~\ref{fig:assert}).

\pagebreak
\section{Summary and Conclusion}\label{sec:conclude}

In this paper, we present the behavior for a simulated individual used in microsimulation of pedestrian crowd. The behavior is based on the SCT and ``field of desire'' hybrid technique in our attempt to provide cognitive behaviors to the simulated individual so that the crowd can exhibit crowd patterns that are observed in real-life, such as that of arching. With this programming, we experimented on the combination effect of $\CROWDSIZE$ and $\EXITWIDTH$ on $\TIMEARCH$, and the arch profile described by~$M$ and~$m$. Here, we found out that the $\TIMEARCH \propto \frac{1}{\CROWDSIZE \times \EXITWIDTH}$, the $M \propto \frac{\CROWDSIZE}{\EXITWIDTH}$, and the $m \propto \CROWDSIZE \times \EXITWIDTH$. These results suggest that, intuitively, the~$\EXITWIDTH$ should be wide enough at higher~$\CROWDSIZE$ so that $\TIMEARCH$ will happen at the least possible time, and at the same time~$M$ will be shorter, and $m$ will be longer (but still $M>m$). We believe, though, that when we find the values for $\CROWDSIZE$ and $\EXITWIDTH$ where $M=m$, then this is the combination when~$\TIMEARCH$ will be the fastest. Thus, we will investigate this hypothesis in the future.

\section{Acknowledgment}
This research effort is under the research program {\bf Multi-agent Simulation of Student Egress from the ICS Mega Lecture Hall Under Normal, Controlled Emergency, and Panic Situations} funded by the {\em Institute of Computer Science, \UPLB}.
\bibliography{arching-phenomenon}

\begin{thebibliography}{39}
\providecommand{\natexlab}[1]{#1}
\providecommand{\url}[1]{\texttt{#1}}
\expandafter\ifx\csname urlstyle\endcsname\relax
  \providecommand{\doi}[1]{doi: #1}\else
  \providecommand{\doi}{doi: \begingroup \urlstyle{rm}\Url}\fi

\bibitem[Allport(1924)]{allport24}
Floyd~Henry Allport.
\newblock \emph{Social Psychology}.
\newblock Houghton Mifflin, Boston, 1924.

\bibitem[Arns(1993)]{arns93}
Thomas Arns.
\newblock Video films of pedestrian crowds, 1993.

\bibitem[Berk(1974)]{berk74}
Richard~A. Berk.
\newblock A gaming approach to crowd behavior.
\newblock \emph{American Sociological Reviews}, 39:\penalty0 355--373, 1974.

\bibitem[Blumer(1939)]{blumer39}
Herbert~G. Blumer.
\newblock Collective behavior.
\newblock In A.M. Lee, editor, \emph{Principles of Sociology}, pages 67--121.
  1939.

\bibitem[Castro and Pabico(2012)]{castro12}
Francisco Enrique Vicente~G. Castro and Jaderick~P. Pabico.
\newblock Microsimulations of arching, clogging, and bursty exit phenomena in
  crowd dynamics.
\newblock In Allan~L. Sioson, editor, \emph{Proceedings of the 10th National
  Conference on Information Technology Education (NCITE 2012)}. Philippine
  Society of IT Educators Foundation, Inc., 2012.

\bibitem[Castro and Pabico(2013)]{castro13}
Francisco Enrique Vicente~G. Castro and Jaderick~P. Pabico.
\newblock Microsimulations of arching, clogging, and bursty exit phenomena in
  crowd dynamics.
\newblock \emph{Philippine Information Technology Journal}, 6\penalty0
  (1):\penalty0 11--16, 2013.

\bibitem[Epstein(1999)]{epstein99}
Joshua~M. Epstein.
\newblock Agent-based computational models and generative social science.
\newblock \emph{Complexity}, 4\penalty0 (5):\penalty0 41--60, 1999.

\bibitem[Festinger(1954)]{festinger54}
Leon Festinger.
\newblock A theory of social comparison processes.
\newblock \emph{Human Relations}, 7:\penalty0 117--140, 1954.

\bibitem[Franklin and Graesser(1996)]{frankling96}
Stan Franklin and Art Graesser.
\newblock Is it an agent, or just a program? {A} taxonomy for autonomous
  agents.
\newblock In \emph{Proceedings of the Third International Workshop on Agent
  Theories, Architectures, and Languages}. Springer-Verlag, 1996.

\bibitem[Fridman and Kaminka(2010)]{fridman10}
Natalie Fridman and Gal~A. Kaminka.
\newblock Modeling pedestrian crowd behavior based on a cognitive model of
  social comparison theory.
\newblock \emph{Computational and Mathematical Organization Theory},
  16:\penalty0 348--372, 2010.

\bibitem[Ganem(1998)]{ganem98}
Joseph Ganem.
\newblock A behavioral demonstration of {F}ermat\'s principle.
\newblock \emph{The Physics Teacher}, 36, 1998.

\bibitem[Goldenstein et~al.(1999)Goldenstein, Large, and Metaxas]{goldstein99}
Siome Goldenstein, Edward Large, and Dimitris Metaxas.
\newblock Non-linear dynamical system approach to behavior modeling.
\newblock \emph{The Visual Computer}, 15\penalty0 (7):\penalty0 349--364, 1999.

\bibitem[Goldenstein et~al.(2001)Goldenstein, Karavelas, Metaxas, Guibas,
  Aaron, and Goswami]{goldstein01}
Siome Goldenstein, Menelaos Karavelas, Dimitris Metaxas, Leonidas Guibas, Eric
  Aaron, and Ambarish Goswami.
\newblock Scalable nonlinear dynamical systems for agent steering and crowd
  simulation.
\newblock \emph{Computers and Graphics}, 25\penalty0 (6):\penalty0 983--998,
  2001.

\bibitem[Guo and Huang(2008)]{gou08}
R.Y. Guo and H.J. Huang.
\newblock A mobile lattice gas model for simulating pedestrian evacuation.
\newblock \emph{Physica A: Statistical Mechanics and its Applications},
  387\penalty0 (2--3):\penalty0 580--586, 2008.

\bibitem[Helbing(2006)]{helbing06}
Dirk Helbing.
\newblock \emph{Safety Management at Large Events: {T}he Problem of Crowd
  Panic}.
\newblock Institute for Transport and Economics, Dresden University of
  Technology, 2006.

\bibitem[Helbing and Molnar(1995)]{helbing95}
Dirk Helbing and Peter Molnar.
\newblock Social force model for pedestrian dynamics.
\newblock \emph{Physical Review E: Journal for Research in Statistical,
  Nonlinear and Soft-matter Physics}, 51\penalty0 (5):\penalty0 4282--4286,
  1995.

\bibitem[Helbing et~al.(2000)Helbing, Farkas, and Vicsek]{helbing00}
Dirk Helbing, Illes~J. Farkas, and Tamas Vicsek.
\newblock Simulating dynamical features of escape panic.
\newblock \emph{Nature}, 407\penalty0 (6803):\penalty0 487--490, 2000.

\bibitem[Helbing et~al.(2001)Helbing, Molnar, Farkas, and Bolay]{helbing01}
Dirk Helbing, Peter Molnar, Illes~J. Farkas, and Kai Bolay.
\newblock Self-organizing pedestrian movement.
\newblock \emph{Environment and Planning B: Planning and Design}, 28:\penalty0
  361--383, 2001.

\bibitem[Henein and White(2005)]{henein05}
Colin~M. Henein and Tony White.
\newblock Agent-based modeling of forces in crowds.
\newblock In \emph{Multi-agent and Multi-agent-based Simulation: Lecture Notes
  in Computer Science}, volume 3415, pages 173--184. 2005.

\bibitem[Huang et~al.(2009)Huang, Wong, Zhang, Shu, and Lam]{huang09}
Ling Huang, S.C. Wong, Mengping Zhang, Chi-Wang Shu, and {William H.K.} Lam.
\newblock Revisiting hughes' dynamic continuum model for pedestrian flow and
  the development of an efficient solution algorithm.
\newblock \emph{Transportation Research Part B: Methodological}, 43\penalty0
  (1):\penalty0 127--141, 2009.

\bibitem[Hughes(2002)]{hughes02}
Roger~L. Hughes.
\newblock A continuum theory for the flow of pedestrians.
\newblock \emph{Transportation Research Part B: Methodological}, 36\penalty0
  (6):\penalty0 507--535, 2002.

\bibitem[Kapadia et~al.(2009)Kapadia, Singh, Hewlett, and Faloutsos]{kapadia09}
Mubbasir Kapadia, Shawn Singh, Billy Hewlett, and Petros Faloutsos.
\newblock Egocentric affordance fields in pedestrian steering.
\newblock In \emph{Proceedings of the 2009 Symposium on Interactive 3D Graphics
  and Games}, pages 215--223, 2009.

\bibitem[{Le Bon}(1895)]{lebon1895}
Gustave {Le Bon}.
\newblock \emph{The Crowd: A Study of the Popular Mind}.
\newblock N.S. Berg, Dunwoody, Georgia, USA, 1968 edition, 1895.

\bibitem[Lerner et~al.(2009)Lerner, Fitusi, Chrysanthou, and
  Cohen-Or]{lerner09}
Alon Lerner, Eitan Fitusi, Yiorgos Chrysanthou, and Daniel Cohen-Or.
\newblock Fitting behaviors to pedestrian simulations.
\newblock In \emph{Proceedings of the 2009 ACM SIGGRAPH/Eurographics Symposium
  on Computer Animation}, pages 199--208, 2009.

\bibitem[Macal and North(2010)]{macal10}
Charles~M. Macal and Michael~J. North.
\newblock Tutorial on agent-based modelling and simulation.
\newblock \emph{Journal Of Simulation}, 4:\penalty0 151--162, 2010.

\bibitem[Muramatsu et~al.(1999)Muramatsu, Irie, and Nagatani]{marumatsu99}
Masakuni Muramatsu, Tunemasa Irie, and Takashi Nagatani.
\newblock Jamming transition in pedestrian counter flow.
\newblock \emph{Physica A: Statical Mechanics and its Applications},
  267\penalty0 (3--4):\penalty0 487--498, 1999.

\bibitem[Musse and Thalmann(2001)]{musse01}
Soraia~R. Musse and Daniel Thalmann.
\newblock Hierarchical model for real time simulation of virtual human crowds.
\newblock \emph{IEEE Transactions on Visualization and Computer Graphics},
  7\penalty0 (2):\penalty0 152--164, 2001.

\bibitem[Musse et~al.(2007)Musse, Jung, Julio C.S.~Jacques, and Braun]{musse07}
Soraia~R. Musse, Claudio~R. Jung, Jr. Julio C.S.~Jacques, and Adriana Braun.
\newblock Using computer vision to simulate the motion of virtual agents.
\newblock \emph{Computer Animation and Virtual Worlds}, 18\penalty0
  (2):\penalty0 83--93, 2007.

\bibitem[Ngoho and Pabico(2009)]{ngoho09}
Louie Vincent~A. Ngoho and Jaderick~P. Pabico.
\newblock Capturing the dynamics of pedestrian traffic using a machine vision
  system.
\newblock \emph{Philippine Information Technology Journal}, 2\penalty0
  (2):\penalty0 1--11, 2009.

\bibitem[Perez et~al.(2002)Perez, Tapang, Lim, and Saloma]{perez02}
Gay~Jane Perez, Giovanni Tapang, May Lim, and Caesar Saloma.
\newblock Streaming, disruptive interference and power-law behavior in the exit
  dynamics of confined pedestrians.
\newblock \emph{Physica A: Statistical Mechanics and its Applications},
  312\penalty0 (3--4):\penalty0 609--618, 2002.

\bibitem[Reynolds(1987)]{reynolds87}
Craig Reynolds.
\newblock Flocks, herds and schools: {A} distributed behavioral model.
\newblock \emph{Computer Graphics (SIGGRAPH)}, 21\penalty0 (4):\penalty0
  25--34, 1987.

\bibitem[Tajima and Nagatani(2001)]{tajima01}
Yusuke Tajima and Takashi Nagatani.
\newblock Scaling behavior of crowd flow outside a hall.
\newblock \emph{Physica A: Statistical Mechanics and its Applications},
  292\penalty0 (1--4):\penalty0 445--554, 2001.

\bibitem[Teknomo(2011)]{teknomo11}
Kardi Teknomo.
\newblock Multi-agent mesoscopic pedestrian simulation.
\newblock In Henry Adorna and Allan Sioson, editors, \emph{Proceedings of the
  11th Philippine Computing Science Congress}, 2011.

\bibitem[Torrens(2004)]{torrens04}
Paul Morrison~Kevin Torrens.
\newblock \emph{Simulating Sprawl: {A} Dynamic Entity-Based Approach to
  Modelling {N}orth {A}merican Suburban Sprawl Using Cellular Automata and
  Multi-Agent Systems}.
\newblock PhD thesis, University College London, London, 2004.

\bibitem[Treuille et~al.(2006)Treuille, Cooper, and Popovi]{treuille06}
Adrien Treuille, Seth Cooper, and Zoran Popovi.
\newblock Continuum crowds.
\newblock In \emph{Proceedings of the Annual Conference on Computer Graphics
  and Interactive Techniques (SIGGRAPH 06)}, pages 1160--1168, 2006.

\bibitem[Tu and Terzopoulos(1994)]{tu94}
Xiaoyuan Tu and Demetri Terzopoulos.
\newblock Artificial fishes: {P}hysics, locomotion, perception, behavior.
\newblock In \emph{Proceedings of the 21st Annual Conference on Computer
  Graphics and Interactive Techniques (SIGGRAPH 94)}, pages 43--50, 1994.

\bibitem[Varas et~al.(2007)Varas, Cornejo, Mainemer, Toledo, Rogan, Munoz, and
  Valdivia]{varas07}
A.~Varas, M.~Cornejo, D.~Mainemer, B.~Toledo, J.~Rogan, V.~Munoz, and
  J.~Valdivia.
\newblock Cellular automaton model for evacuation process with obstacles.
\newblock \emph{Physica A: Statistical Mechanics and its Applications},
  382\penalty0 (2):\penalty0 631--642, 2007.

\bibitem[Weidmann(1993)]{weidmann93}
U.~Weidmann.
\newblock Transportation technique for pedestrians.
\newblock In \emph{Publication Series of the Institute of Transportation,
  Traffic Engineering, Roads and Railways}, number~90. ETH Zurich, Switzerland,
  1993.
\newblock Translated from German.

\bibitem[Wooldridge and Jennings(1995)]{wooldridge95}
Michael Wooldridge and Nicholas~R. Jennings.
\newblock Intelligent agents: {T}heory and practice.
\newblock \emph{Knowledge Engineering Review}, 10\penalty0 (2):\penalty0
  115--152, 1995.

\end{thebibliography}
\bibliographystyle{plainnat}

\end{document}